\documentclass[journal=jpclcd,manuscript=article]{achemso}
\usepackage[version=3]{mhchem}
\usepackage[T1]{fontenc} 

\usepackage[colorlinks=true,allcolors=blue]{hyperref}
\usepackage{graphicx}
\usepackage{amsmath}
\usepackage{bm}
\usepackage[T1]{fontenc}
\usepackage[latin9]{inputenc}
\usepackage{textcomp}
\usepackage{amstext}
\usepackage[Symbol]{upgreek}
\usepackage{booktabs}
\usepackage{dcolumn}
\usepackage{color}
\usepackage{enumerate}
\usepackage{latexsym,bm}
\usepackage{setspace}

\title{Role of Interlayer Coupling on the Evolution of Band-Edges in Few-Layer Phosphorene}
\author{V. Wang}
\affiliation{Department of Applied Physics, Xi'an University of Technology, Xi'an 710054, China}  
\email{wangvei@icloud.com.}
\author{Y. C. Liu}
\affiliation{Department of Applied Physics, Xi'an Jiaotong University, Xi'an 710049, China}
\alsoaffiliation{Department of Applied Physics, Xi'an University of Technology, Xi'an 710054, China}  
\author{Y. Kawazoe}
\affiliation{New Industry Creation Hatchery Center, Tohoku University, Sendai, Miyagi 980-8579, Japan} 
\alsoaffiliation{Kutateladze Institute of Thermophysics, Siberian Branch of Russian Academy of Sciences, Novosibirsk 630090, Russia}
\author{W. T. Geng}
\affiliation{School of Materials Science \& Engineering, University of Science and Technology Beijing, Beijing 100083, China}
\email{geng@ustb.edu.cn.}

\begin{document}
\begin{abstract}
Using first-principles calculations, we have investigated the evolution of band-edges in few-layer phosphorene as a function of the number of P layers. 
Our results predict that monolayer phosphorene is an indirect band gap semiconductor and its valence band edge is extremely sensitive to strain. Its band gap could undergo an indirect-to-direct transition under a lattice expansion as small as 1\% along zigzag direction. A semi-empirical interlayer coupling model is proposed, which can well reproduce the evolution of valence band-edges obtained by first-principles calculations. We conclude that the interlayer coupling plays a dominated role in the evolution of the band-edges via decreasing both band gap and carrier effective masses with the increase of phosphorene thickness. A scrutiny of the orbital-decomposed band structure provides a better understanding of the upward shift of valence band maximum surpassing that of conduction band minimum.
\begin{tocentry} 
\includegraphics[scale=0.29]{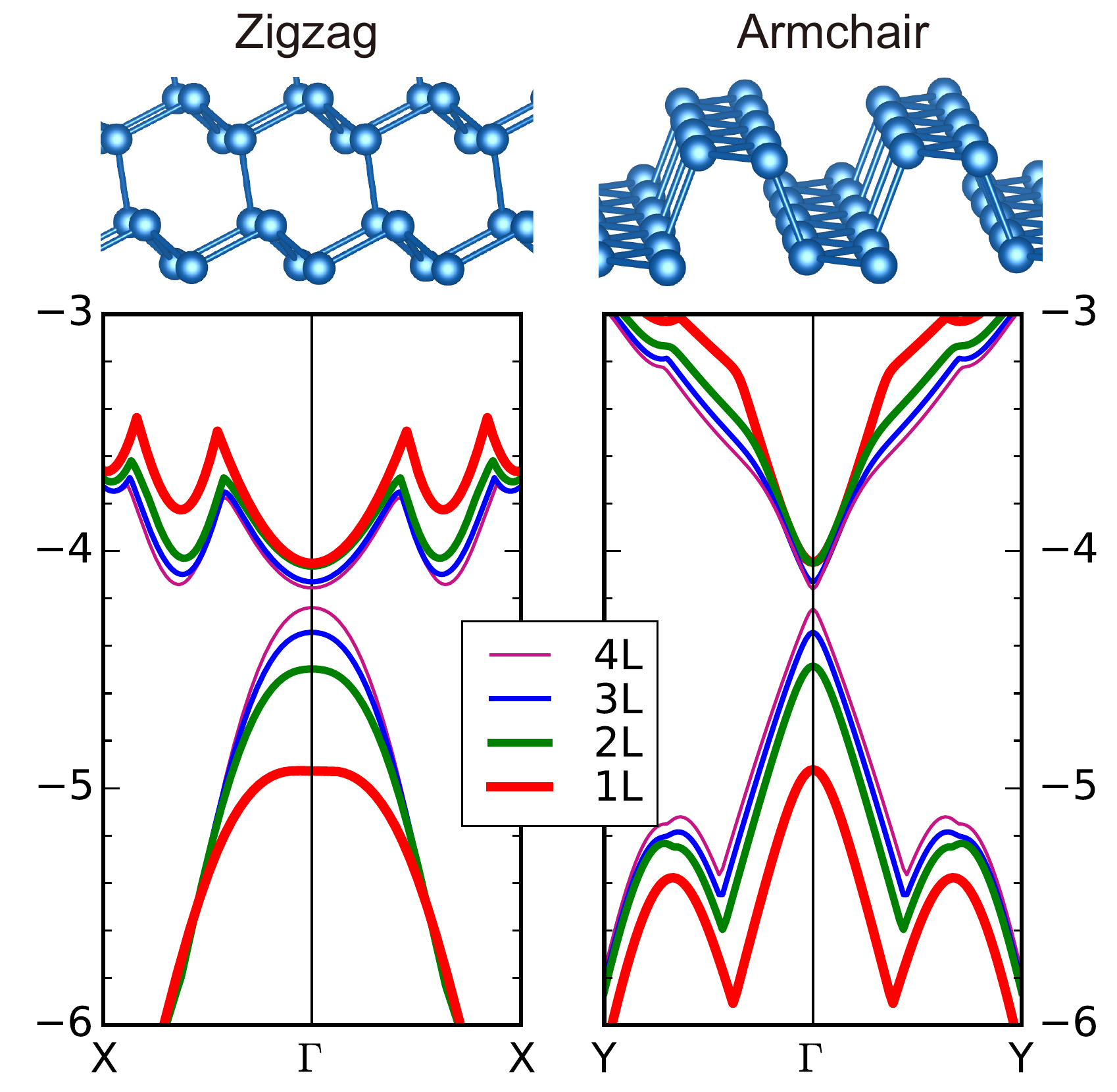}
\end{tocentry}
\end{abstract}
\maketitle 
\vspace{2cm}

\noindent Recently, a novel two-dimensional material, few-layer black phosphorus (phosphorene) has been synthesized by micro-mechanical exfoliation technique.\cite{Li2014,Liu2014} It arouses great interest of researchers 
due to many unparalleled properties superior to or not found in other members of the 2D materials family.\cite{Li2014, Dai2014, Liu2014} For example, it exhibits a carrier mobility up to 1000 cm$^{2}$/V$\cdot$s and an on/off ratio up to 10$^4\sim10^5$ for the phosphorene transistors at room temperature.\cite{Li2014,Liu2014,Koenig2014} Its peculiar puckered honeycomb structure leads to significant anisotropic electronic and optical properties on zigzag and armchair directions.\cite{Ling2015,Qiao2014,Wang2015,Yuan2015} Remarkably, the band gap of phosphorene is thickness-dependent, varying from 0.3 eV in the bulk limit to $\sim$2.2 eV in a monolayer with a direct band gap character.\cite{Keyes1953,Brown1965,Maruyama1981,Akahama1983,Asahina1984,Li2014,Liu2014,Zhang2014,Tran2014,Peng2014,Liu2015,Wang2015,Xia2014} 

Although the structural, electronic and optical properties of few-layer phosphorene have been extensively studied both experimentally and theoretically,\cite{Li2014,Liu2014,Zhang2014,Tran2014,Peng2014,Liu2015,Wang2015,Xia2014,Low2014,Low2014a} 
there are issues that remain controversial or even unexplored, to the best of our knowledge.
For example, there is a debate on whether single-layer phosphorene could have a direct-band gap.\cite{Rodin2014,Li2014a,Rudenko2015,Ganesan2015,Ziletti2015}
Cai \emph{et al.} attributed the thickness-dependence of band gap to the quantum confinement effect. Nevertheless, they noted that monolayer phosphorene is an exception.\cite{Cai2014}
Tran \emph{et al.} predicted that the decay of the band gap of phosphorene with increasing thickness is significantly slower than the usual quantum confinement result.\cite{Tran2014}
This suggests that the quantum confinement effect alone is not enough to describe well the evolution of band gap as a function of the number of P layer.\cite{Rodin2014,Li2014a,Rudenko2015,Ganesan2015,Ziletti2015}
Another unresolved puzzle is that the upward shift of the valence band maximum (VBM) is faster than that of the conduction band minimum (CBM).\cite{Cai2014,wang2015a} 
Obviously, our understanding of the thickness effect on the evolution of the band-edges in phosphorene is still incomplete.

In this Letter, we attempt to answer these questions by examining the fine structure of band edges in few-layer phosphorene using first-principles calculations based on the density functional theory (DFT).\cite{Hohenberg1964,Kohn1965} Our results demonstrate that monolayer phosphorene has an indirect band-gap and 
its valence band edge (VBE) is rather sensitive to the strain. A strain as small as 1\% along zigzag direction is enough to induce an indirect-to-direct band gap transition in monolayer phosphorene. Based on our first-principles results, we  propose a semi-empirical interlayer coupling (SEIC) model to interpret the evolution of the valence band-edges as a function of the P number of layers in few-layer phosphorene. This simple model can reproduce fairly well the evolution of valence band-edges determined by first-principles calculations, a strong indication that interlayer interaction is primarily responsible for the evolution of the band-edges in few-layer phosphorene. 

Our total energy and electronic structure calculations were performed using the Vienna ab initio simulation package (VASP).\cite{Kresse1996, Kresse1996a} The electron-ion interaction was described using projector augmented wave (PAW) method \cite{PAW, Kresse1999} and the exchange and correlation were treated with generalized gradient approximation (GGA) in the Perdew Burke Ernzerhof (PBE) form\cite{Perdew1996}.
It is well known that the GGA method underestimates the semiconductor band gaps. However, in this study, we focus on the evolution of band-edges in few-layer phosphorene; the band dispersions calculated by GGA method exhibit similar features with the exception of the relative position of the valence and conduction bands when comparing with the other higher-level methods, such as hybrid DFT or GW. 
We used a cutoff energy of 400 eV for the plane wave basis set, which yields total energies convergence better than 1 meV/atom. Previous DFT calculations have shown that the interlayer van der Waals (vdW) interaction need to be considered  for a proper description of the geometrical properties of black phosphorus.\cite{Appalakondaiah2012,wang2015a} We therefore incorporated the vdW interactions by employing a semi-empirical correction scheme of Grimme's DFT-D2 method in the following calculations unless otherwise stated, which has been successful in describing the geometries of various layered materials.\cite{Grimme2006, Bucko2010}

For the single-layer phosphorene, we also carried out local density approximation (LDA) using the Ceperley-Alder
exchange correlation potential as parametrized by Perdew and Zunger, \cite{Perdew1981} quasiparticle G0W0 and hybrid DFT \cite{Perdew1996a,Paier2006, Marsman2008} calculations for the purpose of comparison. To achieve good convergence of dielectric function in the G0W0 calculations, we used a large number of 320 bands for the unit cell of monolayer. The converged eigenvalues and wavefunctions, as well as the equilibrium geometry obtained from PBE functional, were chosen as the initial input for the G0W0 calculations. Note that in the G0W0 calculations only the quasiparticle energies were recalculated self-consistently in one iteration; the wave-functions were not updated but remain fixed at the PBE level. 
For visualization purpose, the G0W0 band structure were interpolated to a finer grid using an
interpolation based on Wannier orbitals as implemented in the WANNIER90 code.\cite{Mostofi2008}
In the HSE06 approach, we here employed a revised scheme as proposed by Heyd, Scuseria, and Ernzerhof (HSE06).\cite{Heyd2003,Krukau2006} 
Additional computational settings and a more detailed discussion of the ground-state properties were given in our previous DFT study of the native defect properties in phosphorene.\cite{wang2015a} 

In the slab model of few-layer phosphorene, periodic slabs were separated by a vacuum layer of 15 {\AA} in the \emph{c} direction to avoid mirror interactions. A 8\texttimes{}6\texttimes{}1 \emph{k}-mesh including $\Gamma$-point, generated according to the Monkhorst-Pack scheme,\cite{Monkhorst1976} was applied to the Brillouin-zone integrations. On geometry optimization, both the shapes and internal structural parameters of pristine unit-cells were fully relaxed until the residual force on each atom is less than 0.01 eV/{\AA}. The fine structures of band structures were calculated by sampling 301 \emph{k}-point along each high symmetry line in reciprocal space. The band structures and band edges were aligned with respect to the vacuum level which was determined by aligning the planar-averaged electrostatic potential in the vacuum region far from phosphorene layer. This provides a deeper intuitional insight into the physical origins of band edge evaluation. We used the VASPKIT code for post-processing of the VASP calculated data.\cite{vaspkit}

\begin{figure}[htbp]
\centering
\includegraphics[scale=0.5]{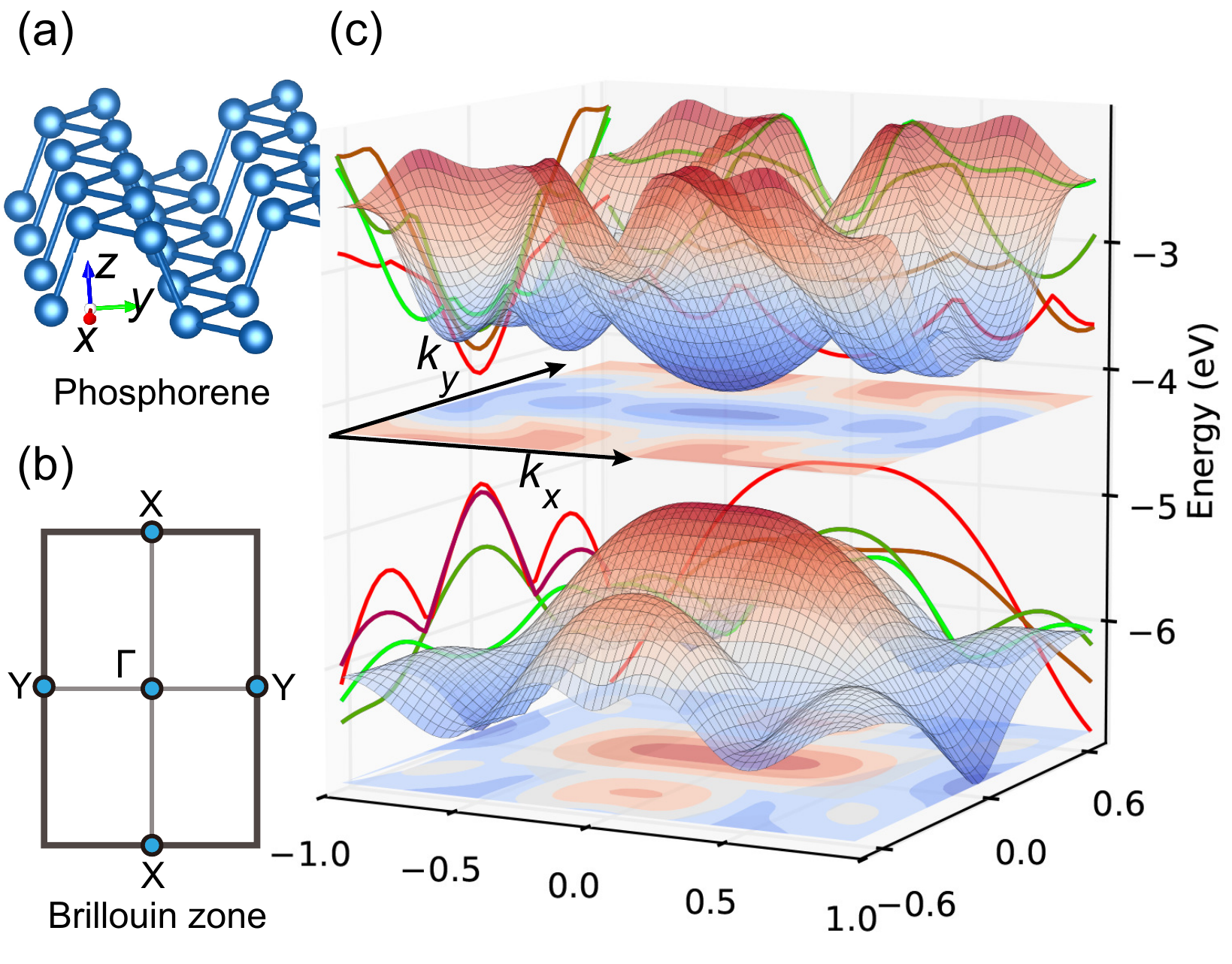}
\caption{\label{3d_band}(Color online) (a) Layered structure, (b) two-dimensional Brillouin zone, and (c) the highest valence and lowest conduction bands of monolayer phosphorene. Green and red curves are the projections of global band edges of monolayer onto the \emph{xz} and \emph{yz} planes. The energy level of vacuum is set to zero.}
\end{figure}

We begin by investigating the global band structure of monolayer phosphorene, aiming to find the locations of both VBM and CBM in the two-dimensional reciprocal lattice space. In Fig. \ref{3d_band}, we find that the band dispersion of VBE along the wavevector \emph{k$_y$} (corresponding to \emph{armchair} direction) near the high symmetry $\Gamma$ point is rather significant, implying a small effective mass of the hole-carrier. 
By contrast, the band dispersion of VBE along the wavevector \emph{k$_x$} (\emph{zigzag} direction) is very flat, a signature of very heavy hole. A similar feature is also observed for the conduction band edge (CBE) dispersion. This means that highly anisotropic effective mass for both electron and hole carriers would be observed in monolayer phosphorene.\cite{Qiao2014} Meanwhile, it appears that both VBM and CBM are located at the $\Gamma$ point in the Brillouin zone, seemingly indicating that the monolayer has a direct band gap.

\begin{figure}[htbp]
\centering
\includegraphics[scale=0.54]{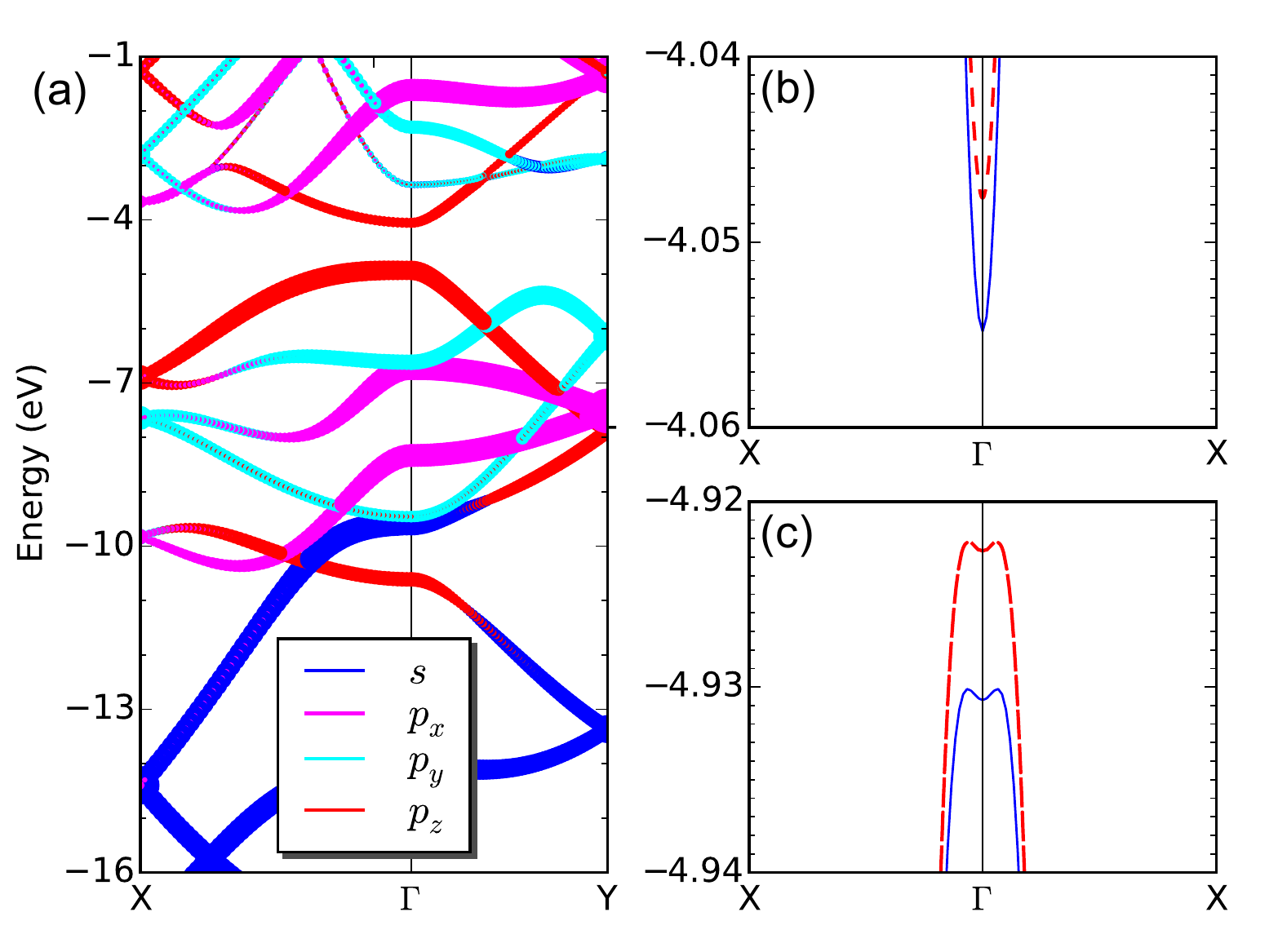}
\caption{\label{monolayer}(Color online) Orbital-projected band structure of monolayer phosphorene (a), and fine structures of CBE (b) and VBE (c) around the $\Gamma$ point. The width of line in (a) indicates the weight of component. Blue solid and red dashed lines in (b) and (c) represent the band edges with and without spin-orbital coupling respectively. The vacuum level is set to zero.}
\end{figure}

To gain further insights into the underlying physics, the orbital-projected band structure of P atom (``fat bands'') are shown in Fig. \ref{monolayer}. The band gap for monolayer phosphorene is predicted to be 0.91 eV by using PBE approach. 
The lowest-lying valence bands are essentially \emph{s} states, characterized by large dispersion along $\Gamma$-X direction. Since the variation of \emph{k$_x$} has marginal effect on the \emph{p}$_y$ orbital, the bands consisting mainly of \emph{p}$_y$ states are relatively flat along $\Gamma$-X direction. The opposite is true for the  bands with \emph{p}$_x$ character in $\Gamma$-Y direction. It is worth mentioning here that the \emph{p}$_x$ and \emph{p}$_y$ orbitals are not orthogonal to each other after hybridization. The VBE are mainly derived from the $\sigma$-$\sigma$ bonding states between \emph{p}$_z$  orbitals (weight percentage: 90.7\%) in different P sublayer. For the CBE, the contributions of \emph{s} (14.8\%) and \emph{p}$_y$ (19.6\%) become comparable with \emph{p}$_z$ (65.6\%). Clearly, the interlayer interaction has a stronger effect on the \emph{p}$_z$  orbitals than on others. It is therefore expected that the position of VBE is more sensitive to the number of P layers than that of CBE due to the interaction of \emph{p}$_z$ between neighboring P layers.

A closer look  of CBE and VBE are displayed in panels (b) and (c) respectively. It is seen clearly that for monolayer phosphorene the CBM is exactly located at $\Gamma$ point, but the  VBM shifts slightly away from $\Gamma$
point by around $\pm$6.33$\times$10$^{-2}$ {\AA}$^{-1}$ along the $\Gamma$-X direction. This means
that monolayer phosphorene is not a direct-band gap semiconductor.  This special position is labeled as $\Lambda$  point
and its energy and position deviation from $\Gamma$ point are denoted as $\Delta$\emph{E}$_{\Lambda}$ and $\Delta$\emph{k}$_{\Lambda}$ respectively. Based on the \textbf{k} $\cdot$ \textbf{p} perturbation theory, Li \emph{et al.} pointed that the indirect band gap character of monolayer phosphorene
originates from the coupling among \emph{s}, \emph{p}$_y$ and \emph{p}$_z$  orbitals since they have the same symmetry of $\Gamma_{2}^{+}$.\cite{Li2014a} Our calculations show that  the inclusion of spin-orbital coupling (SOC) shifts the VBE (CBE) toward (away from) Fermi level by around 8 meV (7 meV). However, the relevant physics discussed above is not affected qualitatively. Thus, the SOC effect will not be considered in the following discussion.

For the purpose of comparison, 
we summarize in Table \ref{offset} the values of $\Delta$\emph{E}$_{\Lambda}$ and $\Delta$\emph{k}$_{\Lambda}$ calculated using various methods with or without DFT-D2 correction. All methods predict that the monolayer phosphorene is not an exact direct gap semiconductor, in accordance with a recent GW work.\cite{Ganesan2015}
Most noticeably of all, the magnitude of $\Delta$\emph{E}$_{\Lambda}$ yielded by HSE06-D2 method even reaches $\sim$100 meV. We remind the reader that the calculated values of $\Delta$\emph{E}$_{\Lambda}$ listed in Table \ref{offset} correspond to different equilibrium lattice constants. 
Further HSE06-D2 calculations using the optimal lattice constants obtained with PBE-D2 (LDA) give $\Delta$\emph{E}$_{\Lambda}$=60 (93) meV and $\Delta$\emph{k}$_{\Lambda}$=$\pm$0.15 (0.16) {\AA}$^{-1}$ respectively. Clearly, the magnitude of $\Delta$\emph{E}$_{\Lambda}$ is sensitive to the P-P bond length. Since the PBE-D2 method still overestimates the lattice constants for bulk black phosphorus,\cite{wang2015a} we speculate that the overestimation may extend to few-layer phosphorene. This might be a non-negligible factor to the very tiny value of $\Delta$\emph{E}$_{\Lambda}$. Moreover, LDA or GGA severely underestimates the semiconductor band gaps, the HSE06 method is expected to yield more accurate $\Delta$\emph{E}$_{\Lambda}$ for given lattice constants. In other words, the PBE method underestimates the energy difference between $\Lambda$ and $\Gamma$ when compared with HSE06 result. 
Interestingly, we find that with a vacuum thickness of 25 {\AA}, G0W0 method gives the values of $\Delta$\emph{E}$_{\Lambda}$ and $\Delta$\emph{k}$_{\Lambda}$ very close to those of LDA. It is, however, worth mentioning that the G0W0 calculated band gap are dependent on vacuum thickness as we shall demonstrate below.
Unfortunately, no experimental data are available for comparison with the predicted values at present.

\begin{table}
  \caption{The calculated lattice constants \emph{a} ({\AA}, zigzag direction) and \emph{b} ({\AA}, armchair direction),  P-P bond length \emph{d} ({\AA}), energy and location deviation $\Delta$\emph{E}$_{\Lambda}$ (meV) and $\Delta$\emph{k}$_{\Lambda}$ ({\AA}$^{-1}$) of $\Lambda$ point away from $\Gamma$ point in monolayer phosphorene, using various methods combined with/without DFT-D2 method.}
  \label{offset}
  \begin{tabular}{llllllll}
    \hline
    XC method & \emph{a}  &  \emph{b} & \emph{d} &$\Delta$\emph{E}$_{\Lambda}$  &  $\Delta$\emph{k}$_{\Lambda}$ \\
    \hline
LDA &3.27  & 4.38 &  2.20$\sim$2.23 &	12 & $\pm$0.13  \\
PBE    & 3.30  &  4.62 & 2.22$\sim$2.26 & 0.6 & $\pm$0.06 \\
PBE-D2 &3.30   &  4.57 & 2.22$\sim$2.25 &  0.7 & $\pm$0.06 \\
HSE06-D2 &3.27 & 4.53 & 2.19$\sim$2.22 &  	99 & $\pm$0.19 \\
G0W0 &3.30   &  4.57 & 2.22$\sim$2.25 &  11 & $\pm$0.12 \\
    \hline
  \end{tabular}
\end{table}

\begin{figure}[htbp]
\centering
\includegraphics[scale=0.54]{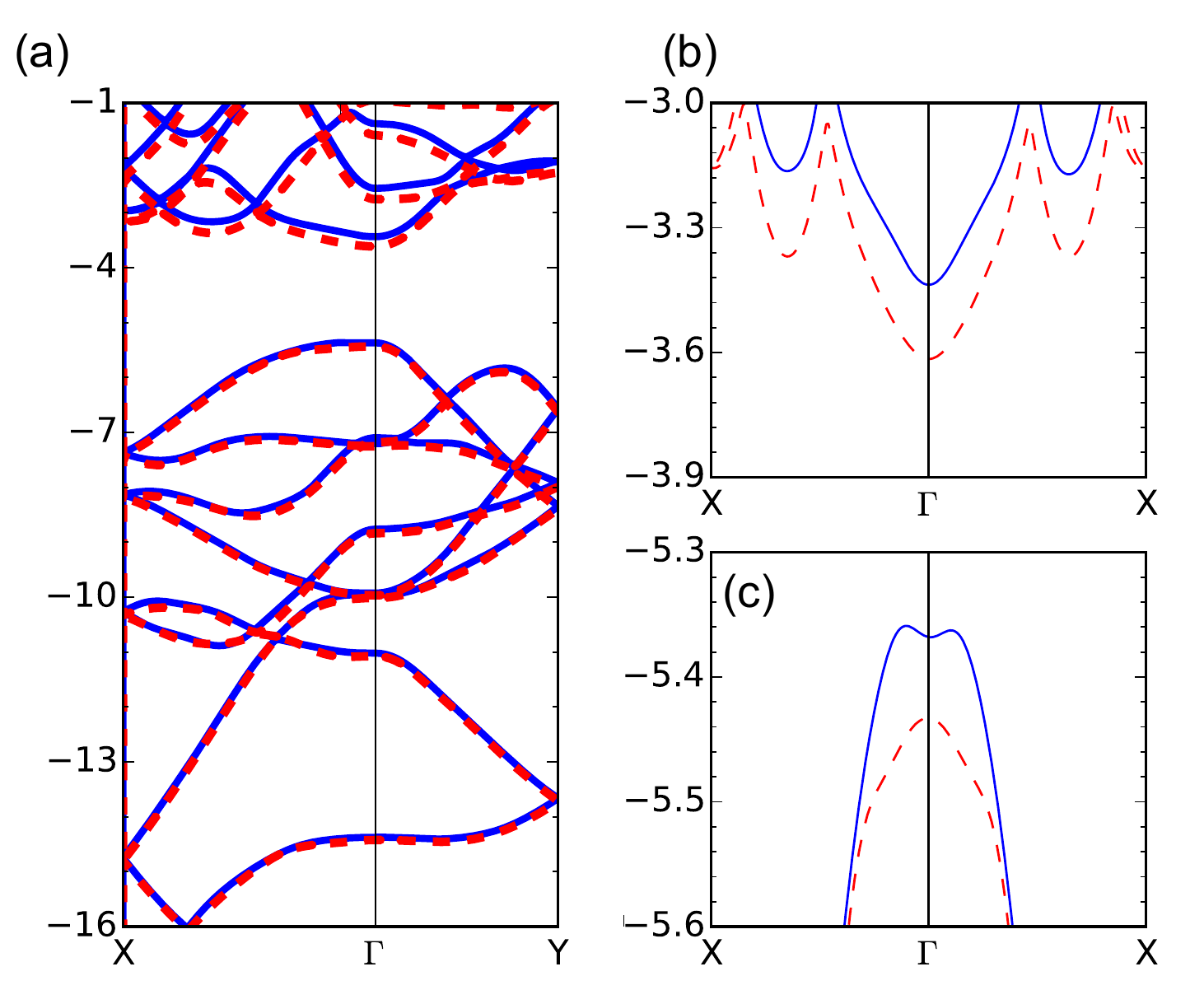}
\caption{\label{g0w0}(Color online) Band structure of monolayer phosphorene given by G0W0 (a) and fine structures of CBE (b) and VBE (c) around the $\Gamma$ point. Blue and red lines represent the results of a slab model with a vacuum thickness of 25 {\AA} and 15 {\AA} respectively. The vacuum level is set to zero.}
\end{figure}

\begin{figure}[htbp]
\centering
\includegraphics[scale=0.5]{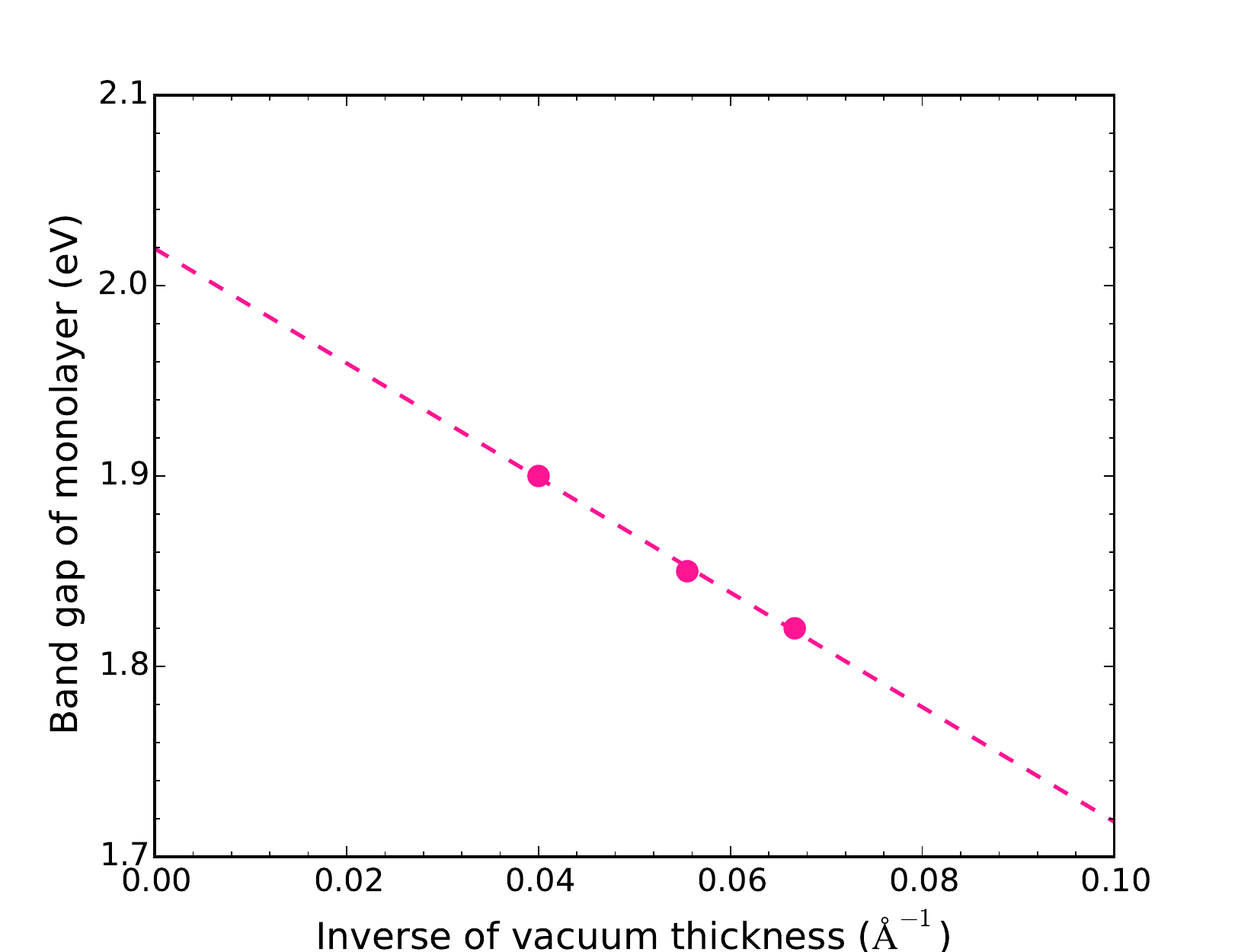}
\caption{\label{g0w0-bandgap}(Color online) Band structure of monolayer phosphorene as a function of the inverse of vacuum thickness given by G0W0.}
\end{figure}

In view of the fact that GW calculations performed by different groups yield inconsistent band character of monolayer phophorene,\cite{Rudenko2015,Ganesan2015,Ziletti2015} we also performed G0W0 calculations to examine the band structure of monolayer as a function of vacuum thickness. We find that the band structure character of monolayer phosphorene, including both band dispersion and band gap, is rather sensitive to the vacuum thickness. As seen in Fig. 3 (a), the band dispersions of monolayer configurations with different vacuum thickness (25 {\AA} versus 15 {\AA}) are qualitatively similar. However, they have different band characters when we look into the fine structure of VBE in Fig. \ref{g0w0}(c). More specifically, with a vacuum thickness of 25 {\AA}, G0W0 predicts an indirect band gap, in agreement with our PBE and HSE06 results. If we adopt a vacuum with a thickness of 15 {\AA} in the slab model, G0W0 predicts a direct band gap. This sensitivity might be the reason for the discrepancy in previous G0W0 results reported in literature. In addition, the gap value also varies linearly as a function of the inverse of vacuum thickness (see Fig. \ref{g0w0-bandgap}). Similar behavior was also found in other two dimensional materials,\cite{Komsa2012,Bhandari2015} reflecting the long-range nature of the self-energy $\Sigma$=\emph{iGW}.\cite{Bhandari2015} Finally, the asymptotic gap value of 2.01 eV for monolayer phosphorene in the limiting case of infinite vacuum is obtained by an extrapolation scheme, very close to the recently reported experimental value of 2.2 eV.\cite{Wang2015}

Next, we investigate the influence of lattice constants on the $\Delta$\emph{E}$_{\Lambda}$ by applying uniform out-of-plane ($\varepsilon_{z}$) or in-plane strain to the unit cell of mononlayer. The latter includes three cases, namely, the uniaxial strain along \emph{x} (zigzag) or \emph{y} (armchair), and the biaxial strain along both \emph{x} and \emph{y} directions, denoted by $\varepsilon_{x}$, $\varepsilon_{y}$ and $\varepsilon_{xy}$ respectively (see Fig. \ref{3d_band}). Taking the case of the uniaxial strain along zigzag direction as an example, the applied strain is defined as $\varepsilon_{x}$=$\frac{a_{x}-a_{x0}}{a_{x0}}$, where \emph{a}$_x$ and \emph{a}$_{x0}$ are the lattice constants along the \emph{x} direction for the strained and relaxed structures. 
A positive (negative) value of $\varepsilon$ corresponds to a tensile (compressive) strain.

\begin{figure}[htbp]
\centering
\includegraphics[scale=0.35]{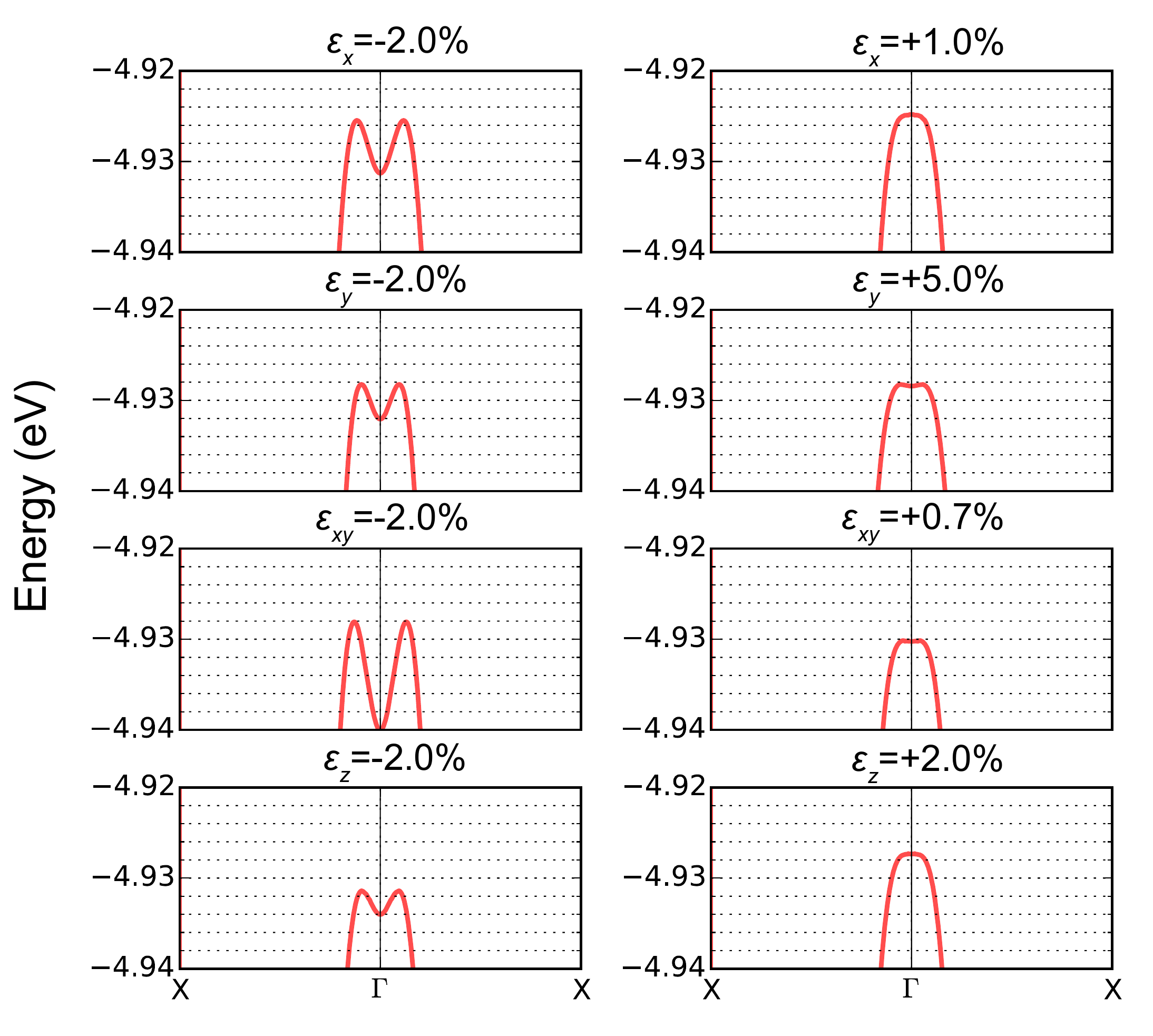}
\caption{\label{strain}(Color online) 
Evolution of VBE for monolayer phosphorene in response to strains determined by PBE-D2. For explanation of $\varepsilon_{x}$, $\varepsilon_{y}$, $\varepsilon_{z}$ and $\varepsilon_{xy}$, please see text.}
\end{figure}

\begin{figure}[htbp]
\centering
\includegraphics[scale=0.4]{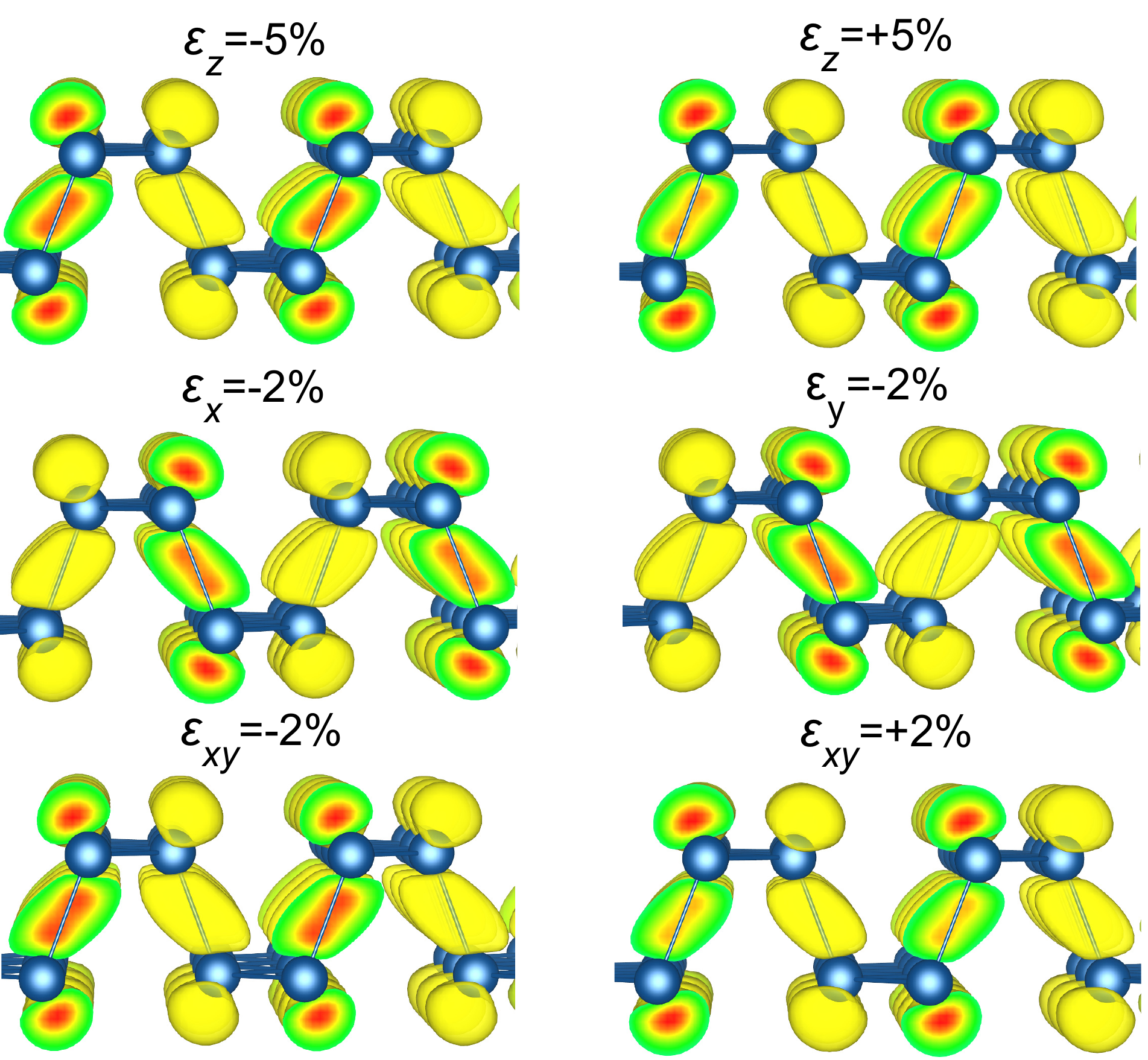}
\caption{\label{partchg}(Color online) 
Charge density plots of the $\Gamma$ point when monolayer phosphorene is subject to varies strains. The isosurface is 0.01 \emph{e}/Bohr$^3$.} 
\end{figure}

As seen in Fig. \ref{strain}, a uniform out-of-plane or in-plane compression lowers the eigenvalue at the $\Gamma$ point. This leads to more significant indirect-gap character of monolayer phosphorene. In contrast, the system undergoes a transition from indirect to direct band gap when tensile strains are applied. Our calculations predict that a small tensile strain of around 1\% along zigzag direction will induce such transition. This implies that the band gap character of thin phosphorene films grown on different substrate could be different due to the introduction of strain by lattice mismatch. 
In contrast, the evolution of VBE is not sensitively dependent on the strain along armchair direction. It is readily understandable if we consider the fact that the VBM is located at the $\Lambda$ point in $\Gamma$-X path (corresponding to zigzag direction) as pointed out above. Previous first-principles calculations also showed that both the electronic and optical properties of single-layer phosphorene depend strongly on the applied strain.\cite{Ciakfmmodemathlseiir2014,Seixas2015}
To gain insight into such observed trend in the VBE of monolayer phosphorene, we examine the bonding characters of $\Gamma$ point by plotting the band-decomposed electron density in Fig. \ref{partchg}.

We find that the depth of concave-like shape of VBE near $\Gamma$ point is associated with the strength of \emph{p}\emph{p}$\sigma$ bonding states between \emph{p}$_z$ orbitals of P atoms. More specifically, stronger bonding strength between \emph{p}$_z$ orbitals induces more significant indirect band gap character.  Bonding strength can be enhanced by reducing the P-P bond length in different sublayers, \emph{i.e.}, applying a compressive out-of-plane or in-plane strain. Phosphorus atoms have five electrons on 3\emph{p} orbitals, three of which participate in the formation of three covalent $\sigma$-bonds with three neighboring P atoms by \emph{sp}$^3$ hybridization in a puckered honeycomb structure, and the remaining two electrons occupying a lone pair orbital oriented out-of-plane. Note that the magnitude of $\Delta$\emph{E}$_{\Lambda}$ is more sensitive to in-plane strain than out-of-plane one. This means that the indirect-gap character of VBE is also likely associated with the coulomb repulsion between electrons from the lone pair orbitals of neighboring P atoms.

\begin{figure}[htbp]
\centering
\includegraphics[scale=0.54]{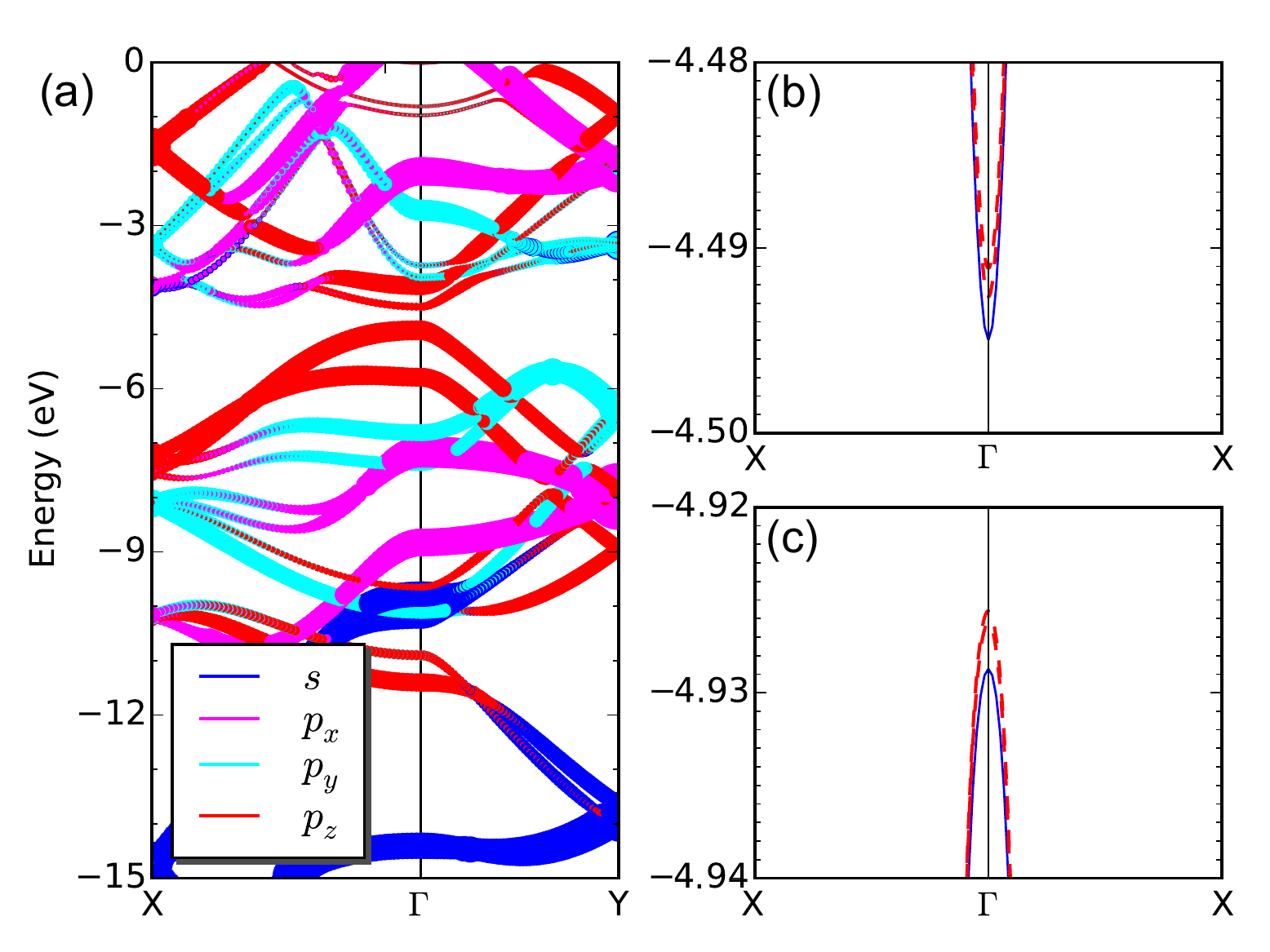}
\caption{\label{bilayer}(Color online) Orbital-projected band structure of bilayer phosphorene (a) and fine structures of CBE (b) and VBE (c) around the $\Gamma$ point. The width of line in (a) indicates the weight of component. Blue solid and red dashed lines in (b) and (c) represent the band edges with and without spin-orbital coupling respectively. The vacuum level is set to zero.}
\end{figure}

When two or more isolated monolayers move close to each other to form multilayer phosphorene, the corresponding degenerated energy levels of different monolayers become non-degenerated due to the interlayer coupling, as is shown in Fig. \ref{bilayer}. This leads to the splitting of the corresponding bands, which in turn push VBM and CBM to higher and lower energies respectively. As a result, the magnitude of band gap decreases with increasing the number of layers.  
On the basis of the careful analysis of the fine structure of conduction and valence bands, as shown in panels (b) and (c), we conclude that bilayer phosphorene is a direct gap semiconductor. In other words, a transition from indirect to direct band gap occurres when going from monolayer to bilayer.

\begin{figure}[htbp]
\centering
\includegraphics[scale=0.38]{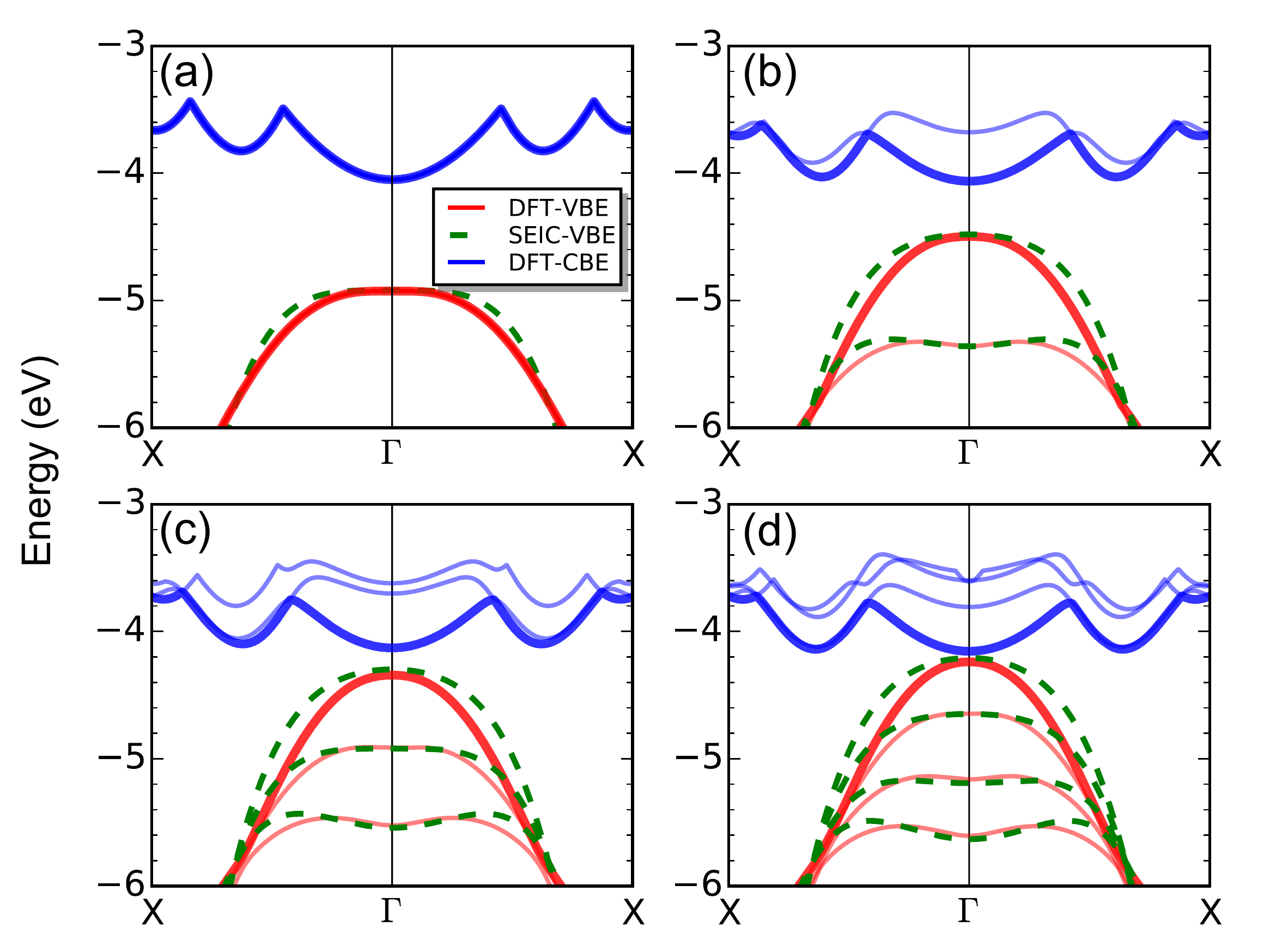}
\caption{\label{band_kx}(Color online) The evolution of the band-edges along $\Gamma$-X direction as a function of the number of layers calculated by DFT and SEIC model. (a) monolayer, (b) bilayer, (c) trilayer, (d) quadrilayer. The vacuum level is set to zero.}
\end{figure}

\begin{figure}[htbp]
\centering
\includegraphics[scale=0.38]{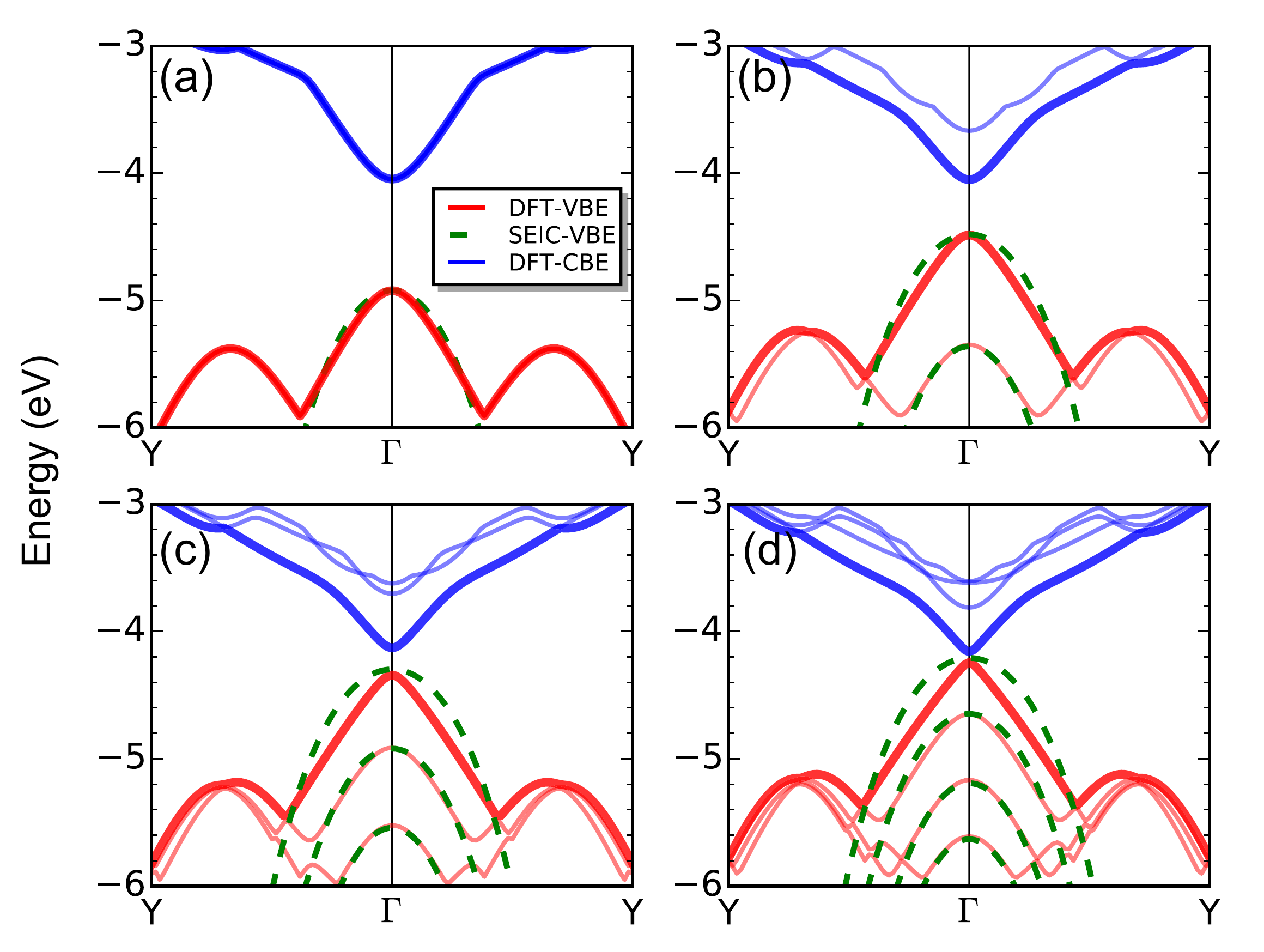}
\caption{\label{band_ky}(Color online) The evolution of the band-edges along $\Gamma$-Y direction as a function of the number of layers calculated by DFT and SEIC model. (a) monolayer, (b) bilayer, (c) trilayer, (d) quadrilayer. The vacuum level is set to zero.}
\end{figure}

Figures \ref{band_kx} and \ref{band_ky} show the evolution of band-edges along $\Gamma$-X and $\Gamma$-Y paths as a function of the number of P layers respectively. One can find the band dispersion of VBE along $\Gamma$-X path becomes more significant with increasing film thickness due to the strong effect of interlayer coupling on \emph{p}$_z$ orbital, suggesting that a significant decrease of hole-carrier mass is to be observed with increasing the number of layers. Generally, a decrease in the carrier effective mass means an enhanced mobility. Therefore, few-layer phosphorene is likely to have better performance than monolayer phosphorene in electronic devices. 

We find that the band dispersions of both VBE and CBE along $\Gamma$-Y path change from parabolic-like shape for monolayer to (nearly) linear-like shape for quadrilayer, suggesting that the Dirac-point observed in graphene could be also realized in few-layer phosphorene by tuning its band gap value to zero eV. On the other hand, the VBE along the $\Gamma$-X direction remain parabolic-like shape, indicating a significantly anisotropic band structure of quadrilayer. We note a very recent experimental study has reported that the tunable band gap and anisotropic Dirac semimetal state can be achieved by sprinkling potassium atoms on top of multilayer phosphorene.\cite{Kim2015} 
Several recent theoretical studies also predicted that the Dirac cone in phosphorene can be achieved by application of an external electric field or strain.\cite{Liu2015a,Fei2015,Dolui2015} 

For quadrilayer, it should be pointed that the PBE method predicts a tiny gap value of around 0.1 eV, which is far smaller than the value of 0.71 (1.08) eV given by HSE06 (GW0).\cite{wang2015a} Therefore, we attribute the linear dispersion of quadrilayer to the strong interaction between valence band and conduction band as a result of the severe gap underestimation by PBE. Further HSE06 calculations reveal that the band dispersion of quadrilayer remains a parabolic-like shape but has a tendency toward linear-dispersion when going from monolayer to the bulk limit, as shown in Supporting Information Figure S1.
It is found that the VBM shifts toward Fermi level by 0.7 eV while the CBM shifts downward by 0.1 eV when going from monolayer to quadrilayer. The fact that the weight of \emph{p}$_z$ in VBE is larger than that in CBE is likely to be responsible for such observations. A similar trend has been found for MoS$_2$.\cite{Padilha2014}
The global evolution of both VBE and CBE for bilayer, trilayer and quadrilayer systems is depicted in Supporting Information Figure S2.

It is now clear that the behavior of VBE is crucial for the electronic properties of phosphorene systems. Thus, we take the case of VBE as an example and propose a SEIC model  
to describe the evolution of VBE as a function of the number of P layers. As an extended H$\ddot{\text{u}}$ckel molecular orbital theory,\cite{Yates2012} the SEIC model can describe the interlayer interaction in layered materials. 
In this model, a multi-layer phosphorene is seen as a ``molecule'', a monolayer is taken as an ``atom'', the energy bands in a monolayer are analogue of electron orbitals, and the interlayer coupling is the hybridization of the orbitals originating from different ``atom'' with the same symmetry..
Based on \textbf{k} $\cdot$ \textbf{p} perturbation theory\cite{Li2014a}, 
the band dispersion of VBE near the $\Gamma$ point in monolayer can be approximately expressed as 
\begin{equation}
\begin{split}
E_{1v}(k_x,k_y)=H_{1v}(k_x,k_y)=\varepsilon_{0}+\alpha_x k_x^{2}+\beta_x k_x^{4} \\
+\alpha_y k_y^{2}+\beta_y k_y^{4},
\end{split}
\end{equation}
where $\varepsilon_{0}$=-4.92 eV is the eigenvalue of $\Gamma$ point of monolayer. The parameters $\alpha_x$, $\alpha_y$, $\beta_x$ and $\beta_y$ are 0.04 eV$\cdot ${\AA}$^2$, -30.0 eV$\cdot ${\AA}$^2$, -80.0 eV$\cdot ${\AA}$^4$ and -100.0 eV$\cdot ${\AA}$^4$ respectively, indicating the anisotropic electronic properties on zigzag and armchair directions. These values can be obtained by fitting the DFT calculated band dispersion of VBE for monolayer phosphorene. Here \emph{k}$_x$ and \emph{k}$_y$ are given in units of 2$\pi$/\emph{a} and 2$\pi$/\emph{b} respectively, where \emph{a} and \emph{b} are the corresponding equilibrium lattice constants of few-layer phosphorene systems. Since we focus only on the interpretation for the evolution of the valence band-edges along zigzag (\emph{k}$_y$=0) and armchair (\emph{k}$_x$=0) directions of few-layer phosphorene, the contribution of coupling between \emph{k}$_x$ and \emph{k}$_y$ to the valence band dispersions is neglected. 

As for the multilayer, the interlayer interaction is described by a coupling term of \emph{J}$_v$(\textbf{k}). For simplicity, the interaction between non-neighboring layers is neglected without loss of accuracy. In the case of VBE, the coupling of \emph{p}$_z$-\emph{p}$_z$ orbitals in nearest-neighboring layers is only considered since the VBE is mainly derived from \emph{p}$_z$ state as pointed above. Then the interlayer coupling term of VBE is defined as
\begin{equation}
J_v(k_x, k_y)=\mu+\nu_x k_x^{2}+\nu_y k_y^{2},
\end{equation}
where $\mu$, $\nu_x$ and $\nu_y$ are the coupling parameters.

The Hamiltonian of valence band dispersions near the $\Gamma$ point in bilayer is 
\begin{equation}\label{eq3}
H_{2v}(k_x, k_y)=\left \lgroup
\begin{matrix}
H_{1v}&J_v \\
J_v &H_{1v}
\end{matrix}
\right \rgroup,
\end{equation} 

and the eigenvalues of Eqs. (\ref{eq3}) are 
\begin{equation}\label{eq5}
E_{2v}^{\pm}(k_x, k_y)=E_{1v}(k_x, k_y){\pm}J_v(k_x, k_y).
\end{equation} 

Based on the DFT-calculated results of valence band dispersions for bilayer as displayed in Fig. \ref{band_kx} (b), the parameters $\mu$, $\nu_x$ and $\nu_y$ are 0.44 eV, -4.18 eV$\cdot ${\AA}$^2$ and 6.0 eV$\cdot ${\AA}$^2$ respectively. One can find that the energy of the first and second highest valence band of bilayer are shifted upward and downward respectively by \emph{J}$_v$(\emph{k}$_x$,\emph{k}$_y$) eV, with respect to monolayer. Note that \emph{J}$_v$(\emph{k}$_x$,\emph{k}$_y$) is wavevector-dependent and reaches its maximum $\mu$=0.44 eV at the $\Gamma$ point. 
Clearly, our SEIC model reproduces the indirect-direct band gap transition when going from monolayer to bilayer as predicted by our DFT results.
Next, we take trilayer as an example to study the role of interlayer coupling on the evolution of VBE in multilayer system.  The Hamiltonian of trilayer and the corresponding eigenvalues can be described as

\begin{equation}\label{eq4}
H_{3v}(k_x, k_y)=\left \lgroup
\begin{matrix}
H_{1v}&J_v &0\\
J_v &H_{1v} & J_v\\
0&J_v &H_{1v}\\
\end{matrix}
\right \rgroup,
\end{equation} 
and
\begin{equation}\label{eq5} 
\begin{split}                                                                         
E_{3v}^1(k_x, k_y)=E_{1v}(k_x, k_y)-\sqrt{2}J_v(k_x, k_y), \\                                      
E_{3v}^2(k_x, k_y)=E_{1v}(k_x, k_y), \\                                                   
E_{3v}^3(k_x, k_y)=E_{1v}(k_x, k_y)+\sqrt{2}J_v(k_x, k_y).    
\end{split}                                  
\end{equation}      
                                          
The analysis is similar for the systems with more layers. For the \emph{n}-layer system, the evolution of its VBE is determined by \emph{E}$_{nv}$(\emph{k}$_x$, \emph{k}$_y$)=\emph{E}$_{1v}$+2\emph{J}$_v$$\cdot$cos$\frac{k}{n+1}\pi$, where \emph{k}=1, 2, ..., \emph{n} is the index number of \emph{k}$^{th}$ splitting valence band. In the bulk limit (\emph{n} $\to$ $\infty$), the eigenvalue of $\Gamma$ point is expected to be \emph{E}$_{1v}$+2\emph{J}$_v$=-3.16 eV with respect to the vacuum level. In other words, compared with monolayer the gap value of bulk black phosphorus reduces at least 0.88 eV if one also consider the evolution of CBE toward Fermi level. This prediction agrees well with our previous DFT-PBE results.\cite{wang2015a}

Overall, the SEIC model results are in good agreement with those obtained from DFT, as shown in Figs. \ref{band_kx} and \ref{band_ky}. 
However, the SEIC model predicts that the VBE of quadrilayer or more-layer systems along $\Gamma$-Y direction still remain quadratic-like shape, instead of showing linear-like shape from DFT calculations [see Figs. \ref{band_ky} (c) and (d)]. 
This disagreement is probably caused by the negligence of the stronger coupling between VBE and CBE when they get close in the quadrilayer or thicker systems. 
However, it should be emphasized here that the artificial linear dispersion of quadrilayer is derived from the problem of the band gap underestimation related to the local or semilocal approximations of the exchange correlation functional in DFT.
Based on the first-order \textbf{k} $\cdot$ \textbf{p} perturbation theory,\cite{Dresselhaus2007} we use a degenerate two-band model to describe the coupling between the valence and conduction bands for these systems. Our results indicate that this coupling does lead to the linear-dominated dispersion along armchair direction and the quadratic-dominated dispersion along zigzag direction respectively.

We expect our SEIC model is also applicable to the evolution of CBE. The evolution of CBE is different from that of VBE. As an illustrative example we choose the case of trilayer system, one can observe in Fig. \ref{band_kx} (c) that the energies of the second and third conduction bands near the $\Gamma$ point become very close to each other; while the second and third valence bands is separated by an energy of $\sqrt{2}J_v(k_x, k_y)$ eV. This means the coupling among \emph{s}, \emph{p}$_y$ and \emph{p}$_z$ become noticeable and cannot be neglected. Further calculations show that the relative positions of the first four conduction bands at the $\Gamma$ point can be qualitatively determined if we include the \emph{s} and \emph{p}$_y$ states into the SEIC model to describe the evolution of the CBE.
Previous theoretical studies suggested that the decreasing tendency of band gap with increasing film thickness is likely an outcome of the quantum confinement effect.\cite{Cai2014,Tran2014}  We argue that it would be sufficient to explain such an observed tendency using a simple SEIC model. Clearly, the dramatic decrease of hole mass is also attributed to interlayer coupling. We hope that this systematic study serves as a guideline for the interpretation of the evolution of band edges in other two-dimensional layered materials.

To conclude, through first-principles calculations and a semi-empirical interlayer coupling model, we have uncovered the nature and evolution of the valence band-edge as a function of the number of P layers in few-layer phosphorene. Our results predict that the interlayer coupling play a vital role in determining the decreasing trend for both band gap and carrier effective masses with the increase of phosphorene thickness. Also, the interlayer coupling leads to a transition from indirect to direct when changing from monolayer to few-layer phosphorene. The analysis of the orbital-decomposed band structure reveal that the upward shift of valence band maximum is faster than that of conduction band minimum, due to the larger contribution of \emph{p}$_z$ in the former. 

\begin{acknowledgement}
We acknowledges the financial support of the Special Scientific Research Program of the Education Bureau of Shaanxi Province, China (Grant No. 15JK1531), National Natural Science Foundation of Special Theoretical Physics (Grant No. 11447217) and National Natural Science Foundation of China (Grant Nos. 11304245 and 61308006). Y.K. is thankful to the Russian Megagrant Project No.14.B25.31.0030 ``New energy technologies and energy carriers''
for supporting the present research. The calculations were performed on the HITACHI SR16000 supercomputer at the Institute for Materials Research of Tohoku University, Japan.

\textbf{Supporting Information}: The Supporting Information is available free of charge on the ACS Publications website at DOI: {\color{blue}{10.1021/acs.jpclett.5b02047}}: <HSE06-calculated band edges of quadrilayer phosphorene and bulk black phosphorous along $\Gamma$-Y direction; PBE-calculated global evolution of both VBE and CBE for few-layer quadrilayer phosphorene are given in the Supporting Information respectively.>
\end{acknowledgement}

\nocite{*}
\bibliographystyle{biochem}
\bibliography{references}
\end{document}